\documentclass[letterpaper,twocolumn,prl,aps,showpacs,superscriptaddress]{revtex4}
\usepackage[latin1]{inputenc}
\usepackage{bm}
\usepackage{multirow}
\usepackage{amssymb}
\usepackage{amsbsy}
\usepackage{amsmath}
\usepackage{stmaryrd}
\usepackage{graphicx}
\usepackage{epsfig}
\usepackage{placeins}
\makeatletter

\begin{document}

\title{Projection Operator Approach to \\ Spin Diffusion in the Anisotropic Heisenberg Chain at High Temperatures}

\author{Robin Steinigeweg}%
\email{r.steinigeweg@tu-bs.de}%
\affiliation{Institute for Theoretical Physics, Technical University Braunschweig, Mendelsohnstr.~3, D-38106 Braunschweig, Germany}%

\author{Roman Schnalle}%
\email{rschnall@uni-bielefeld.de}%
\affiliation{Department of Physics, University Bielefeld, Universit\"atsstr.~25, D-33615 Bielefeld, Germany}%

\date{\today}

\begin{abstract}
We investigate spin transport in the anisotropic Heisenberg chain in
the limit of high temperatures ($\beta \rightarrow 0$). We
particularly focus on diffusion and the quantitative evaluation of
diffusion constants from current autocorrelations as a function of
the anisotropy parameter $\Delta$ and the spin quantum number $s$.
Our approach is essentially based on an application of the
time-convolutionless (TCL) projection operator technique. Within
this perturbative approach the projection onto the current yields
the decay of autocorrelations to lowest order of $\Delta$. The
resulting diffusion constants scale as $\Delta^{-2}$ in the
Markovian regime $\Delta \ll 1$ ($s=1/2$) and as $\Delta^{-1}$ in
the highly non-Markovian regime above $\Delta \sim 1$ (arbitrary
$s$). In the latter regime the dependence on $s$ appears
approximately as an overall scaling factor $\sqrt{s(s+1)}$ only.
These results are in remarkably good agreement with diffusion
constants for $\Delta > 1$ which are obtained directly from the
exact diagonalization of autocorrelations or have been obtained from
non-equilibrium bath scenarios.
\end{abstract}



\maketitle

Transport in one-dimensional quantum systems has been a topic of
theoretical investigations since at least the 1960s
\cite{huber1969-1, huber1969-2}. Even nowadays there is an ongoing
and still increasing interest in understanding the transport
phenomena in such systems, including their dependence on length
scales and temperature \cite{zotos1999, benz2005, sirker2009,
heidrichmeisner2003, prelovsek2004, michel2008, prosen2009,
steinigeweg2009-1, steinigeweg2009-2, castella1995, jung2007,
jung2006, karadamoglou2004}. Much work has been devoted to a rather
qualitative classification of the emerging transport types into
either non-normal ballistic or normal diffusive behavior. In this
context the crucial mechanisms for the emergence of pure diffusion
have been intensively studied and in particular non-integrability or
quantum chaos are frequently discussed w.r.t.~their role as an at
least necessary prerequisite \cite{castella1995, jung2007, jung2006}.
Spin chains are a central issue of research with a considerable
focus on spin transport in the anisotropic Heisenberg chain, in
particular on the case $s=1/2$ \cite{zotos1999, benz2005,
sirker2009, heidrichmeisner2003, prelovsek2004, michel2008,
prosen2009, steinigeweg2009-1, steinigeweg2009-2, castella1995,
jung2007}. In that case there still is an unsettled debate about the
concrete range of anisotropies where the dynamics at finite
temperatures is indeed diffusive \cite{zotos1999, benz2005,
sirker2009}.
\\
However, even if the diffusive range of anisotropies was definitely
known, the question about the quantitative value of the diffusion
coefficient arises naturally. In fact, this question is as
challenging as the proof of diffusion as such. Because the majority
of all available approaches to transport has to deal more or less
with finite quantum systems, signatures of diffusion in the
thermodynamic limit may not be observed, e.g., due to large mean
free paths, even in the limit of high temperatures. For these
temperatures diffusion coefficients can be found in the literature
for $s=1$ and the isotropic point \cite{karadamoglou2004,
huber1969-1, huber1969-2} and for $s=1/2$ and a narrow window of
large anisotropies above the isotropic point \cite{prelovsek2004,
michel2008, prosen2009, steinigeweg2009-1, steinigeweg2009-2}
solely. On that account there is an urgent need for a much more
comprehensive picture of diffusion constants.
\\
In this work we intend to make an essential step towards such a
picture at high temperatures. To this end we focus on the
quantitative evaluation of diffusion constants from current
autocorrelations, either directly or perturbatively by the use of
projection operator techniques, see Refs.
\onlinecite{chaturvedi1979, breuer2007,nakajima1958, zwanzig1960}.
Within this perturbative approach the projection onto the current
yields the decay of autocorrelations to lowest order of the
anisotropy parameter. The resulting diffusion constant is a smooth
function of the anisotropy and agrees well with the above ones from
the literature for large anisotropies, i.e., at the upper boundary
of the expected range of validity. This agreement also supports the
validity of the perturbative approach in the limit of small
anisotropies, if transport is indeed diffusive.

The anisotropic Heisenberg chain (XXZ model) may be described by a
Hamiltonian of the form $\hat{H} = \hat{H}_0 + \Delta \, \hat{V}$,
where $\hat{H_0}$ and $\hat{V}$ are given by
\begin{equation}
\hat{H}_0 = J \sum_{\mu = 1}^{N} \hat{s}_\mu^x \hat{s}_{\mu+1}^x +
\hat{s}_\mu^y \hat{s}_{\mu+1}^y \, , \quad \hat{V} = J
\sum_{\mu=1}^N \hat{s}_\mu^z \hat{s}_{\mu+1}^z
\end{equation}
with periodic boundary conditions. Here, $J$ denotes the nearest
neighbor coupling strength, $\Delta$ the anisotropy, $N$ the number
of sites, and the operators $\hat{s}_\mu^i$ ($i = x,y,z$) represent
the standard spin matrices (w.r.t.~site $\mu$) for the spin quantum
number $s$. For the clarity of later notations we define the
abbreviation $\tilde{s} \equiv \sqrt{s(s+1)}$.

The Einstein relation $\sigma_\text{dc} = \chi \, {\cal D}$ connects
the spin (dc) conductivity $\sigma_\text{dc}$ and the spin diffusion
constant $\cal D$ via the static susceptibility $\chi$. Therefore
the spin diffusion constant according to linear response theory
({\bf LR}) \cite{kubo1991} reads in the limit of high temperatures
($\beta \rightarrow 0$)
\begin{equation}
{\cal D} = \lim_{t \rightarrow \infty} {\cal D}(t) \, , \quad {\cal
D}(t) = \frac{\beta}{\chi \, \text{dim} {\cal H}  \, N} \int_0^t
\text{d}t' \, C(t') \, , \label{D}
\end{equation}
where $\chi = 1/3 \, \beta \, \tilde{s}^2$, $\text{dim} {\cal H} =
(2s+1)^N$, and the spin current autocorrelation function is given by
\begin{equation}
C(t) = \text{Tr} \{ \hat{J}(t) \hat{J}(0) \} \, , \quad \hat{J} = J
\sum_{\mu = 1}^{N} \hat{s}_\mu^x \hat{s}_{\mu+1}^y - \hat{s}_\mu^y
\hat{s}_{\mu+1}^x
\end{equation}
with the initial value $\text{Tr} \{ \hat{J}^2 \} = 2/9 \, J^2 \,
\tilde{s}^4 \, \text{dim}{\cal H} \, N$. For $s=1/2$ the infinite
time integral in Eq.~(\ref{D}) diverges for $\Delta = 0$ due to $[
\hat{H}_0, \hat{J} ] = 0$ \cite{note}. It also diverges, whenever
there is only partial conservation, i.e., a non-zero Drude weight.
Finite Drude weights indicate ballistic transport at the infinite
time scale but are commonly expected to vanish in the thermodynamic
limit for the non-integrable cases $s > 1/2$
\cite{karadamoglou2004}. For the integrable case $s=1/2$ Drude
weights are widely expected to vanish for $\Delta > 1$
\cite{heidrichmeisner2003, prelovsek2004}, while they may already be
zero for $\Delta = 1$ \cite{zotos1999, benz2005} or even for all $0
< \Delta < 1$ \cite{sirker2009}.
\\
Apart from the common {\bf LR} picture, the time-dependent diffusion
coefficient ${\cal D}(t)$ according to Eq.~(\ref{D}) can be also
connected to the time evolution of the spatial variance of an
initially inhomogeneous non-equilibrium density, see
Refs.~\onlinecite{steinigeweg2009-1, steinigeweg2009-2}. Within this
picture the density dynamics is diffusive at a certain time,
respectively length scale, if ${\cal D}(t)$ is constant at that
scale. Such a scale may exist, even if ${\cal D}(t)$ eventually
diverges in the infinite time limit, i.e., despite finite Drude
weights. We thus do not focus only on the infinite time limit, but
investigate the full time dependence of ${\cal D}(t)$, too.

\begin{figure}[htb]
\includegraphics[width=1.0\linewidth]{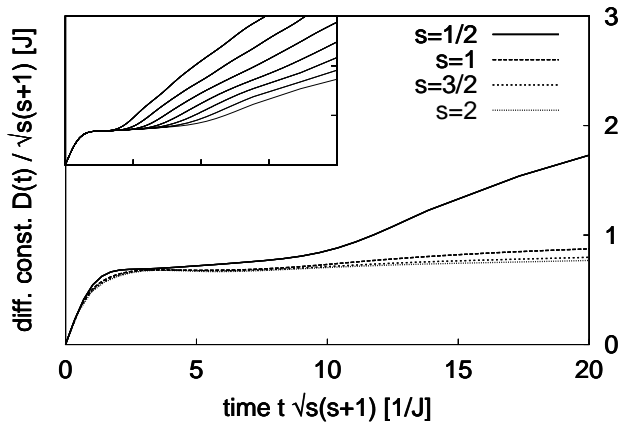}
\caption{The time dependence of the spin diffusion constant ${\cal
D}(t)$ in the limit of high temperatures ($\beta \rightarrow 0$), as
obtained directly from exact diagonalization. The data is displayed
for the anisotropy parameter $\Delta=1.5$ and for the spin quantum
numbers $s=1/2$ ($N=20$), $s=1$ ($N=12$), $s=3/2$ ($N=9$), and $s=2$
($N=8$). The data in the inset is shown for the spin quantum number
$s=1/2$ and for various chain lengths $N = 8, 10, \ldots, 20$.}
\label{figure_ED_D}
\end{figure}

\begin{figure}[htb]
\includegraphics[width=1.0\linewidth]{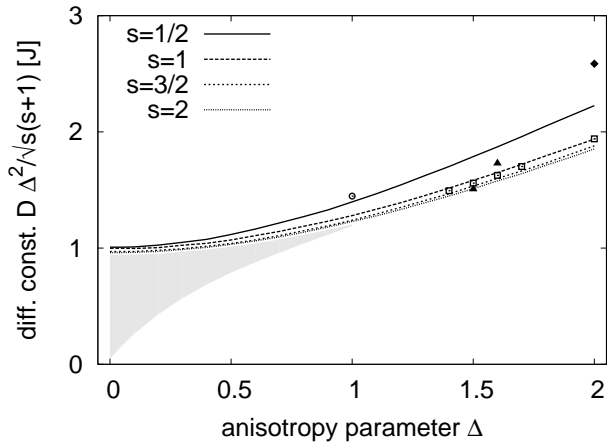}
\caption{The spin diffusion constant ${\cal D}$ as a function of the
anisotropy parameter $\Delta$ in the limit of high temperatures
($\beta \rightarrow 0$), as predicted by TCL to lowest order of
$\Delta$, for the spin quantum numbers $s=1/2, \ldots, 2$ (curves).
Further data from exact diagonalization (squares, $s \geq 1/2$) and
from the literature are indicated: Ref.~\onlinecite{prelovsek2004}
(rhombus, $s = 1/2$), Refs.~\onlinecite{michel2008, prosen2009}
(triangles, $s = 1/2$), and Refs.~\onlinecite{karadamoglou2004,
huber1969-1,huber1969-2} (circle, $s = 1$). Curves for $s > 1/2$ and
$\Delta \ll 1$ have to be understood in terms of an upper boundary.
($\cal D$ is located somewhere below, e.g., as {\it sketched} by the
filled area.)} \label{figure_TCL_D}
\end{figure}

A possible strategy for the analysis of the dependence of ${\cal
D}(t)$ on time is the direct application of numerically exact
diagonalization ({\bf ED}). Of course, {\bf ED} is restricted to a
rather limited range of $N$, even if all symmetries are taken into
account, e.g., translation, rotation about the $z$-axis
\cite{schnalle2009}. But finite $N$ effects are often less
pronounced at finite time scales. In Fig.~\ref{figure_ED_D} we show
${\cal D}(t)$ at such time scales for $\Delta = 1.5$ and the maximum
$N$ which is available to us for a certain $s$: ${\cal D}(t)$
firstly increases at short times $t \lesssim 2/(J \, \tilde{s})$ and
then becomes approximately constant at intermediate times, i.e., a
``plateau'' with a height of about ${\cal D} \approx 0.7 \, J \,
\tilde{s}$ is formed here. ${\cal D}(t)$ finally increases again due
to non-zero Drude weights for finite $N$. But in the thermodynamic
limit Drude weights are expected to vanish for $\Delta = 1.5$ and
each $s$. Furthermore, the height of the plateau does not change
with $N$, while its width gradually increases, see
Fig.~\ref{figure_ED_D} (inset). The height of this plateau
consequently is a reasonable suggestion for the concrete value of
the diffusion coefficient \cite{steinigeweg2009-1,
steinigeweg2009-2}. In fact, for $s=1/2$ the value ${\cal D} \approx
0.6 \, J$ is in remarkably good agreement with non-equilibrium bath
scenarios \cite{michel2008, prosen2009}.
\\
Analogously, the concept of plateaus allows to suggest further
values of the diffusion coefficient for anisotropy parameters above
$\Delta \sim 1.4$, see Fig.~\ref{figure_TCL_D} (squares). Here, the
dependence of $\cal D$ on $s$ appears approximately as a scaling
factor $\tilde{s}$. For anisotropy parameters below $\Delta \sim
1.4$ the suggestion of diffusion coefficients is not possible,
because Drude weights become dominant and plateaus disappear for
finite $N$. Plateaus appear only, if contributions from the Drude
weight and the regular part are well-balanced, at least to some
degree. However, plateaus are also not visible, if both
contributions are treated separately from each other
\cite{steinigeweg2009-2}.

\begin{figure}[htb]
\includegraphics[width=1.0\linewidth]{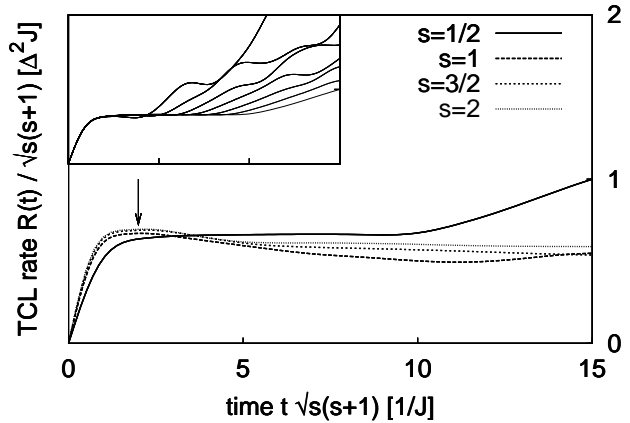}
\caption{ The time dependence of the decay rate $R(t)$, as given
by Eq.~(\ref{TCL_rate}). The data is shown for the spin quantum
numbers $s=1/2$ ($N=20$), $s=1$ ($N=12$), $s=3/2$ ($N=9$), and
$s=2$ ($N=8$). The data in the inset is displayed for the spin
quantum number $s=1/2$ and for various chain lengths $N = 8, 10,
\ldots, 20$. The extracted plateau height for $s > 1/2$ is
indicated (arrow).}
\label{figure_TCL_rate}
\end{figure}

Another strategy for the analysis of the time evolution of $C(t)$,
respectively ${\cal D}(t)$ is provided by an application of the
time-convolutionless ({\bf TCL}) projection operator technique
\cite{chaturvedi1979, breuer2007}. This technique, and the
well-known Nakajima-Zwanzig ({\bf NZ}) method \cite{nakajima1958,
zwanzig1960}, are commonly applied in order to describe the reduced
dynamics for a set of relevant variables. Their application
essentially requires a Hamiltonian of the form $\hat{H} = \hat{H}_0
+ \Delta \, \hat{V}$ and the commutation of the relevant variables
with $\hat{H}_0$ (or a comparatively slow dynamics
w.r.t.~$\hat{H}_0$). Although both methods are well-established
approaches in the context of open quantum systems, the following
approach to current autocorrelation functions in closed quantum
systems is an uncommon concept. However, in this context {\bf NZ} is
very similar to the more common Mori-Zwanzig memory matrix
formalism, see Ref.~\onlinecite{jung2006}. But a ``Mori-TCL''
variant does not exist, of course.
\\
The above-mentioned set of relevant variables is specified by the
definition of a suitable projection (super)operator $\cal P$ which
projects out the relevant part of an operator $\hat{\rho}(t)$. (This
operator does not need to be a density matrix in the strict sense.)
To this end we define
\begin{equation}
\hat{A} \equiv \text{Tr} \{ \hat{J}^2 \}^{-\frac{1}{2}} \, \hat{J}
\, , \quad {\cal P} \, \rho(t) \equiv \text{Tr} \{ \hat{A} \,
\hat{\rho}(t) \} \, \hat{A} \, ,
\end{equation}
where the normalized current operator $\hat{A}$ is introduced in
order to satisfy the property ${\cal P}^2 = {\cal P}$ of a
projection (super)operator.
\\
For the initial condition $\hat{\rho}(0) = \hat{A}$ the {\bf TCL}
technique routinely yields a homogenous differential equation for
the actual expectation value $a(t) \equiv \text{Tr} \{ \hat{A}(t) \,
\hat{A}_\text{I}(t) \}$. Here, the index I of $\hat{A}_\text{I}(t)$
denotes the interaction picture, i.e., the Heisenberg picture
w.r.t.~$\hat{H}_0$. Since $C(t) \propto a(t)$ for $\hat{A}_I(t) =
\hat{A}$ (or $\hat{A}_I(t) \approx \hat{A}$ at a pertinent time
scale \cite{note}), the {\bf TCL} equation is identical for $C(t)$
in that case. It then reads
\begin{equation}
\frac{\text{d}}{\text{d}t} \, C(t) = -R(t) \, C(t) \, , \quad R(t) =
\sum_{i = 1}^\infty \Delta^{2i} \, R_{2i}(t) \label{TCL_equation}
\end{equation}
and avoids the often troublesome time-convolution which appears in
the {\bf NZ} variant \cite{chaturvedi1979, breuer2007}. The
time-dependent rate $R(t)$ is given in terms of a systematic
perturbation expansion in powers of $\Delta$. (All odd orders of
$\Delta$ vanish for this and many other quantum systems.) The
truncation of $R(t)$ to lowest order of $\Delta$ reads
\begin{equation}
R(t) \approx \Delta^2 \int_0^t \text{d}t' \, f(t') \label{TCL_rate}
\end{equation}
with the two-point correlation function
\begin{equation}
f(t) = \text{Tr} \{ \hat{J}^2 \}^{-1} \, \text{Tr} \{ \, \imath
[\hat{J}, \hat{V}]_\text{I}(t) \, \imath [\hat{J},
\hat{V}]_\text{I}(0) \, \} \, ,
\end{equation}
where the above index I indicates again the Heisenberg picture
w.r.t.~$\hat{H}_0$.
In general this lowest order truncation is expected to be justified
for small $\Delta$ (and short $t$). Even though the incorporation of
higher order corrections is in principle possible, their concrete
evaluation is an almost impossible task, both from an analytical and
numerical point of view.
\\
The concrete evaluation of the lowest order truncation for $R(t)$ in
Eq.~(\ref{TCL_rate}) is feasible. Moreover, this evaluation may be
done analytically, if $\hat{H}_0$ can be brought in diagonal form,
e.g., via the Jordan-Wigner transformation onto non-interacting
spinless fermions for $s=1/2$. However, since the expression for
$R(t)$ still takes on a non-trivial form after such a
transformation, it may possibly have to be evaluated numerically at
the end. We thus directly evaluate $R(t)$ in
Fig.~\ref{figure_TCL_rate} numerically by the use of {\bf ED},
similarly as done in Refs.~\onlinecite{jung2007, jung2006}.

There is an apparent similarity between Figs.~\ref{figure_ED_D} and
\ref{figure_TCL_rate}, even though completely different quantities
are shown. In particular a constant rate $R \approx 0.67 \, \Delta^2
\, J \, \tilde{s}$ may be read off from the height of the plateau at
intermediate times, at least for $s = 1/2$. Of course, for $s > 1/2$
the plateau is not developed as clearly for finite $N$. But the
respective $R(t)$-curves have converged completely for short times
$t \lesssim 2/(J \, \tilde{s})$ solely. We hence assume a constant
rate $R$ in that case also, e.g., as given by the height of the
local maximum at such times, see Fig.~\ref{figure_TCL_rate} (arrow).
Obviously, this assumption implies a positive $R$, i.e., we expect
that {\bf TCL} does simply not predict Drude weights to lowest order,
see below. However, in principle $R(t)$ may become zero in the
long-time limit.
\\
The above assumption is relevant only in the Markovian limit of
long relaxation times, i.e., small $\Delta$. In that limit the rate
$R(t)$ can be replaced by the constant value $R$ and lowest order
{\bf TCL} consequently predicts the exponential decay $C(t) = C(0)
\exp(-R \, t)$, cf.~Eq.~(\ref{TCL_equation}). The diffusion constant
in Eq.~(\ref{D}) eventually scales as ${\cal D} \propto R^{-1}
\propto \Delta^{-2}$, e.g., as generally expected from a Boltzmann
equation approach to the same question. The Markovian limit of long
relaxation times concretely turns out to be realized for $\Delta \ll
1$,
see Fig.~\ref{figure_TCL_D}.
\\
However, in that limit the resulting diffusion constants for $s >
1/2$ have to be understood in terms of an upper boundary. Since
$\hat{A}_\text{I}(t) \approx \hat{A}$ does not hold true for long
relaxation times, a back-transform from the interaction picture
becomes necessary in order to obtain $C(t)$ from $a(t)$. Because we
do not perform this back-transform due to its unavailability, we
only consider the contribution from $\hat{V}$ and neglect any
contribution from $\hat{H}_0$, i.e., $C(t)$ probably decays too
slowly. Obviously, $C(t)$ decays too slowly, if exclusively
$\hat{H}_0$ contributes, i.e., for $\Delta = 0$.
\\
For small $\Delta$ there is no substantial difference between the
predictions of lowest order {\bf TCL} and {\bf NZ}. But the
situation changes for large $\Delta$, when relaxation times are
short and non-Markovian effects become relevant, i.e., the initial
increase of the rate $R(t)$ at short times, see
Fig.~\ref{figure_TCL_rate}. At such times the $R(t)$-curve has well
converged already for each $s$ and additional assumptions like the
one above are not required here. Since the initial increase appears
to be approximately linear, lowest order {\bf TCL} predicts a
Gaussian decay in the highly non-Markovian limit. As a consequence
the diffusion constant scales as ${\cal D} \propto \Delta^{-1}$ now.
The highly non-Markovian limit of very short relaxation times turns
out to be concretely realized for anisotropies above $\Delta \sim
1$, see Fig.~\ref{figure_TCL_D}.

For $s>1/2$ the diffusion constants according to lowest order {\bf
TCL} in Fig.~\ref{figure_TCL_D} are in excellent agreement with the
results from {\bf ED} for $\Delta > 1$ in Fig.~\ref{figure_ED_D}.
Furthermore, these diffusion constants are in good agreement with
results for $\Delta = 1$ in the literature (standard frequency
moments analysis, microcanonical Lanczos method
\cite{huber1969-1,huber1969-2,karadamoglou2004}). For $s=1/2$ there
is a deviation on the order of $20 \%$ between diffusion
coefficients according to lowest order {\bf TCL} and results for
$\Delta \sim 1.5$ from {\bf ED} in Fig.~\ref{figure_ED_D} or the
literature (non-equilibrium bath scenarios \cite{michel2008,
prosen2009}). Such a deviation probably arises from the neglection
of all higher order corrections which are more important for
$s=1/2$.
\\
Of course, the above agreement also supports the validity of lowest
order {\bf TCL} for $s=1/2$ and $\Delta \ll 1$, e.g., the usual
range of application for a perturbation theory. But one has to keep
in mind that this perturbation theory is generally restricted to
small $\Delta$ and short $t$ as well. The restriction becomes
manifest, since we expect that {\bf TCL} does simply not predict Drude
weights to lowest order, i.e., the non-decaying character of $C(t)$
essentially is a higher order effect, as known for large $\Delta$.
Consequently, lowest order {\bf TCL} can be valid at all $t$ only,
if Drude weights eventually vanish in the thermodynamic limit.
\\
On that account we may finally conclude as follows: Whenever there
is spin diffusion in the XXZ model at high temperatures, either at a
finite or the infinite time scale, we expect that the respective
diffusion coefficient is reasonably suggested by lowest order {\bf
TCL}, e.g., as displayed for $s=1/2$ in Fig.~\ref{figure_TCL_D}.

\begin{acknowledgments}
We sincerely thank Wolfram Brenig, J\"urgen Schnack, and Jochen
Gemmer for fruitful discussions. We further gratefully acknowledge
financial support by the {\it Deutsche Forschungsgemeinschaft},
one of us (RST) through FOR 912.
\end{acknowledgments}


\begin{thebibliography}{22}
\expandafter\ifx\csname natexlab\endcsname\relax\def\natexlab#1{#1}\fi
\expandafter\ifx\csname bibnamefont\endcsname\relax
  \def\bibnamefont#1{#1}\fi
\expandafter\ifx\csname bibfnamefont\endcsname\relax
  \def\bibfnamefont#1{#1}\fi
\expandafter\ifx\csname citenamefont\endcsname\relax
  \def\citenamefont#1{#1}\fi
\expandafter\ifx\csname url\endcsname\relax
  \def\url#1{\texttt{#1}}\fi
\expandafter\ifx\csname urlprefix\endcsname\relax\def\urlprefix{URL }\fi
\providecommand{\bibinfo}[2]{#2}
\providecommand{\eprint}[2][]{\url{#2}}

\bibitem[{\citenamefont{{D. L. Huber} and {J. S. Semura}}(1969)}]{huber1969-1}
\bibinfo{author}{\bibnamefont{{D. L. Huber}}} \bibnamefont{and}
  \bibinfo{author}{\bibnamefont{{J. S. Semura}}}, \bibinfo{journal}{Phys. Rev.}
  \textbf{\bibinfo{volume}{182}}, \bibinfo{pages}{602} (\bibinfo{year}{1969}).

\bibitem[{\citenamefont{{D. L. Huber} et~al.}(1969)\citenamefont{{D. L. Huber},
  {J. S. Semura}, and {C. G. Windsor}}}]{huber1969-2}
\bibinfo{author}{\bibnamefont{{D. L. Huber}}},
  \bibinfo{author}{\bibnamefont{{J. S. Semura}}}, \bibnamefont{and}
  \bibinfo{author}{\bibnamefont{{C. G. Windsor}}}, \bibinfo{journal}{Phys.
  Rev.} \textbf{\bibinfo{volume}{186}}, \bibinfo{pages}{534}
  (\bibinfo{year}{1969}).

\bibitem[{\citenamefont{{X. Zotos}}(1999)}]{zotos1999}
\bibinfo{author}{\bibnamefont{{X. Zotos}}}, \bibinfo{journal}{Phys. Rev. Lett.}
  \textbf{\bibinfo{volume}{82}}, \bibinfo{pages}{1764} (\bibinfo{year}{1999}).

\bibitem[{\citenamefont{{J. Benz} et~al.}(2005)\citenamefont{{J. Benz}, {T.
  Fukui}, {A. Kl\"umper}, and {C. Scheeren}}}]{benz2005}
\bibinfo{author}{\bibnamefont{{J. Benz}}}, \bibinfo{author}{\bibnamefont{{T.
  Fukui}}}, \bibinfo{author}{\bibnamefont{{A. Kl\"umper}}}, \bibnamefont{and}
  \bibinfo{author}{\bibnamefont{{C. Scheeren}}}, \bibinfo{journal}{J. Phys.
  Soc. Jpn. Suppl.} \textbf{\bibinfo{volume}{74}}, \bibinfo{pages}{181}
  (\bibinfo{year}{2005}).

\bibitem[{\citenamefont{{J. Sirker} et~al.}(2009)\citenamefont{{J. Sirker}, {R.
  G. Pereira}, and {I. Affleck}}}]{sirker2009}
\bibinfo{author}{\bibnamefont{{J. Sirker}}}, \bibinfo{author}{\bibnamefont{{R.
  G. Pereira}}}, \bibnamefont{and} \bibinfo{author}{\bibnamefont{{I.
  Affleck}}}, \bibinfo{journal}{Phys. Rev. Lett.}
  \textbf{\bibinfo{volume}{103}}, \bibinfo{pages}{216602}
  (\bibinfo{year}{2009}).

\bibitem[{\citenamefont{{F. Heidrich-Meisner} et~al.}(2003)\citenamefont{{F.
  Heidrich-Meisner}, {A. Honecker}, {D. C. Cabra}, and {W.
  Brenig}}}]{heidrichmeisner2003}
\bibinfo{author}{\bibnamefont{{F. Heidrich-Meisner}}},
  \bibinfo{author}{\bibnamefont{{A. Honecker}}},
  \bibinfo{author}{\bibnamefont{{D. C. Cabra}}}, \bibnamefont{and}
  \bibinfo{author}{\bibnamefont{{W. Brenig}}}, \bibinfo{journal}{Phys. Rev. B}
  \textbf{\bibinfo{volume}{68}}, \bibinfo{pages}{134436}
  (\bibinfo{year}{2003}).

\bibitem[{\citenamefont{{P. Prelov\v{s}ek} et~al.}(2004)\citenamefont{{P.
  Prelov\v{s}ek}, {S. El Shawish}, {X. Zotos}, and {M. Long}}}]{prelovsek2004}
\bibinfo{author}{\bibnamefont{{P. Prelov\v{s}ek}}},
  \bibinfo{author}{\bibnamefont{{S. El Shawish}}},
  \bibinfo{author}{\bibnamefont{{X. Zotos}}}, \bibnamefont{and}
  \bibinfo{author}{\bibnamefont{{M. Long}}}, \bibinfo{journal}{Phys. Rev. B}
  \textbf{\bibinfo{volume}{70}}, \bibinfo{pages}{205129}
  (\bibinfo{year}{2004}).

\bibitem[{\citenamefont{{M. Michel} et~al.}(2008)\citenamefont{{M. Michel}, {O.
  Hess}, {H. Wichterich}, and {J. Gemmer}}}]{michel2008}
\bibinfo{author}{\bibnamefont{{M. Michel}}}, \bibinfo{author}{\bibnamefont{{O.
  Hess}}}, \bibinfo{author}{\bibnamefont{{H. Wichterich}}}, \bibnamefont{and}
  \bibinfo{author}{\bibnamefont{{J. Gemmer}}}, \bibinfo{journal}{Phys. Rev. B}
  \textbf{\bibinfo{volume}{77}}, \bibinfo{pages}{104303}
  (\bibinfo{year}{2008}).

\bibitem[{\citenamefont{{T. Prosen} and {M.
  \v{Z}nidari\v{c}}}(2009)}]{prosen2009}
\bibinfo{author}{\bibnamefont{{T. Prosen}}} \bibnamefont{and}
  \bibinfo{author}{\bibnamefont{{M. \v{Z}nidari\v{c}}}}, \bibinfo{journal}{J.
  Stat. Mech.} \textbf{\bibinfo{volume}{2009}}, \bibinfo{pages}{P02035}
  (\bibinfo{year}{2009}).

\bibitem[{\citenamefont{{R. Steinigeweg} et~al.}(2009)\citenamefont{{R.
  Steinigeweg}, {H. Wichterich}, and {J. Gemmer}}}]{steinigeweg2009-1}
\bibinfo{author}{\bibnamefont{{R. Steinigeweg}}},
  \bibinfo{author}{\bibnamefont{{H. Wichterich}}}, \bibnamefont{and}
  \bibinfo{author}{\bibnamefont{{J. Gemmer}}}, \bibinfo{journal}{EPL}
  \textbf{\bibinfo{volume}{88}}, \bibinfo{pages}{10004} (\bibinfo{year}{2009}).

\bibitem[{\citenamefont{{R. Steinigeweg} and {J.
  Gemmer}}(2009)}]{steinigeweg2009-2}
\bibinfo{author}{\bibnamefont{{R. Steinigeweg}}} \bibnamefont{and}
  \bibinfo{author}{\bibnamefont{{J. Gemmer}}}, \bibinfo{journal}{Phys. Rev. B}
  \textbf{\bibinfo{volume}{80}}, \bibinfo{pages}{184402}
  (\bibinfo{year}{2009}).

\bibitem[{\citenamefont{{H. Castella} et~al.}(1995)\citenamefont{{H. Castella},
  {X. Zotos}, and {P. Prelov\v{s}ek}}}]{castella1995}
\bibinfo{author}{\bibnamefont{{H. Castella}}},
  \bibinfo{author}{\bibnamefont{{X. Zotos}}}, \bibnamefont{and}
  \bibinfo{author}{\bibnamefont{{P. Prelov\v{s}ek}}}, \bibinfo{journal}{Phys.
  Rev. Lett.} \textbf{\bibinfo{volume}{74}}, \bibinfo{pages}{972}
  (\bibinfo{year}{1995}).

\bibitem[{\citenamefont{{P. Jung} and {A. Rosch}}(2007)}]{jung2007}
\bibinfo{author}{\bibnamefont{{P. Jung}}} \bibnamefont{and}
  \bibinfo{author}{\bibnamefont{{A. Rosch}}}, \bibinfo{journal}{Phys. Rev. B}
  \textbf{\bibinfo{volume}{76}}, \bibinfo{pages}{245108}
  (\bibinfo{year}{2007}).

\bibitem[{\citenamefont{{P. Jung} et~al.}(2006)\citenamefont{{P. Jung}, {R. W.
  Helmes}, and {A. Rosch}}}]{jung2006}
\bibinfo{author}{\bibnamefont{{P. Jung}}}, \bibinfo{author}{\bibnamefont{{R. W.
  Helmes}}}, \bibnamefont{and} \bibinfo{author}{\bibnamefont{{A. Rosch}}},
  \bibinfo{journal}{Phys. Rev. Lett.} \textbf{\bibinfo{volume}{96}},
  \bibinfo{pages}{067202} (\bibinfo{year}{2006}).

\bibitem[{\citenamefont{{J. Karadamoglou} and {X.
  Zotos}}(2004)}]{karadamoglou2004}
\bibinfo{author}{\bibnamefont{{J. Karadamoglou}}} \bibnamefont{and}
  \bibinfo{author}{\bibnamefont{{X. Zotos}}}, \bibinfo{journal}{Phys. Rev.
  Lett.} \textbf{\bibinfo{volume}{93}}, \bibinfo{pages}{177203}
  (\bibinfo{year}{2004}).

\bibitem[{\citenamefont{{S. Chaturvedi} and {F.
  Shibata}}(1979)}]{chaturvedi1979}
\bibinfo{author}{\bibnamefont{{S. Chaturvedi}}} \bibnamefont{and}
  \bibinfo{author}{\bibnamefont{{F. Shibata}}}, \bibinfo{journal}{Z. Phys. B}
  \textbf{\bibinfo{volume}{35}}, \bibinfo{pages}{297} (\bibinfo{year}{1979}).

\bibitem[{\citenamefont{{H.-P. Breuer} and {F.
  Petruccione}}(2007)}]{breuer2007}
\bibinfo{author}{\bibnamefont{{H.-P. Breuer}}} \bibnamefont{and}
  \bibinfo{author}{\bibnamefont{{F. Petruccione}}}, \emph{\bibinfo{title}{The
  {T}heory of {O}pen {Q}uantum {S}ystems}} (\bibinfo{publisher}{Oxford
  University Press, New York}, \bibinfo{year}{2007}).

\bibitem[{\citenamefont{{S. Nakajima}}(1958)}]{nakajima1958}
\bibinfo{author}{\bibnamefont{{S. Nakajima}}}, \bibinfo{journal}{Progr. Theor.
  Phys.} \textbf{\bibinfo{volume}{20}}, \bibinfo{pages}{948}
  (\bibinfo{year}{1958}).

\bibitem[{\citenamefont{{R. Zwanzig}}(1960)}]{zwanzig1960}
\bibinfo{author}{\bibnamefont{{R. Zwanzig}}}, \bibinfo{journal}{J. Chem. Phys.}
  \textbf{\bibinfo{volume}{33}}, \bibinfo{pages}{1338} (\bibinfo{year}{1960}).

\bibitem[{\citenamefont{{R. Kubo} et~al.}(1991)\citenamefont{{R. Kubo}, {M.
  Yokota}, and {S. Hashtisume}}}]{kubo1991}
\bibinfo{author}{\bibnamefont{{R. Kubo}}}, \bibinfo{author}{\bibnamefont{{M.
  Yokota}}}, \bibnamefont{and} \bibinfo{author}{\bibnamefont{{S. Hashtisume}}},
  \emph{\bibinfo{title}{Statistical {P}hysics {II}: {N}onequilibrium
  {S}tatistical {M}echanics}}, vol.~\bibinfo{volume}{31} of
  \emph{\bibinfo{series}{Solid State Sciences}} (\bibinfo{publisher}{Springer,
  New York}, \bibinfo{year}{1991}), \bibinfo{edition}{2nd} ed.

\bibitem[{not()}]{note}
\bibinfo{note}{$[ \hat{H}_0, \hat{J} ] \propto \sum_\mu (\hat{s}_\mu^z)^2
  (\hat{s}_{\mu+1}^z - \hat{s}_{\mu-1}^z)$ is non-zero for $s>1/2$ and numerics
  also indicates a (partial) decay of $C(t)$ at a time scale $t \gg 2/( J \,
  \tilde{s})$.}

\bibitem[{\citenamefont{{R. Schnalle} and {J. Schnack}}(2009)}]{schnalle2009}
\bibinfo{author}{\bibnamefont{{R. Schnalle}}} \bibnamefont{and}
  \bibinfo{author}{\bibnamefont{{J. Schnack}}}, \bibinfo{journal}{Phys. Rev. B}
  \textbf{\bibinfo{volume}{79}}, \bibinfo{pages}{104419}
  (\bibinfo{year}{2009}).

\end{thebibliography}

\newpage

\end{document}